\documentclass[final,5p,twocolumn,times,sort&compress]{elsarticle}

\usepackage{graphicx}
\usepackage{graphicx}
\usepackage{amsmath,bm}
\usepackage{flushend}
\usepackage{color}

\def\lsim{\mathrel{\rlap{\lower4pt\hbox{\hskip1pt$\sim$}}
    \raise1pt\hbox{$<$}}}
\def\gsim{\mathrel{\rlap{\lower4pt\hbox{\hskip1pt$\sim$}}
    \raise1pt\hbox{$>$}}}
\def\sqr#1#2{{\vcenter{\vbox{\hrule height.#2pt
         \hbox{\vrule width.#2pt height#1pt \kern#1pt
         \vrule width.#2pt}
         \hrule height.#2pt}}}}
\newcommand{\beq}{\begin{equation}}
\newcommand{\eeq}{\end{equation}}
\newcommand{\bea}{\begin{eqnarray}}
\newcommand{\eea}{\end{eqnarray}}
\newcommand{\rf}[1]{(\ref{#1})}

\begin{document}

\begin{frontmatter}

\title{Classical-physics applications for Finsler $b$ space}

\author{Joshua Foster$^a$ and Ralf Lehnert$^b$}

\address{$^a$Physics Department, Indiana University,
Bloomington, IN 47405, USA\\
$^b$Indiana University Center for Spacetime Symmetries, Bloomington, IN 47405, USA
}

\begin{abstract}

The classical propagation of 
certain Lorentz-violating fermions 
is known to be governed by 
geodesics of a four-dimensional pseudo-Finsler $b$ space 
parametrized by a prescribed background covector field. 
This work identifies systems in classical physics 
that are governed by 
the three-dimensional version of Finsler $b$ space  
and constructs a geodesic 
for a sample non-constant choice 
for the background covector.
The existence of these classical analogues 
demonstrates that 
Finsler $b$ spaces possess applications 
in conventional physics, 
which may yield insight into 
the propagation of SME fermions on curved manifolds.
\end{abstract}
\end{frontmatter}


\section{Introduction}
\label{intro}

In Einstein's general relativity, 
the mathematical description of the classical gravitational field 
is intrinsically geometric  
and rests on the concept of a Riemannian manifold. 
However, 
in the quest for 
a consistent quantum theory of gravitation
this phenomenologically successful and elegant geometric formalism 
may need to be generalized.
A popular broader geometric framework that
incorporates Riemannian manifolds as a special limit
is provided by Riemann--Finsler geometry~\cite{br,pf,bcs,sczs}. 
The recent interest of the physics community 
in studying such geometries~\cite{gyb,Girelli07,Gibbons07,lph,jsmv,VAK_Finsler,sv,V11,rsv,pw11,js,nm,dcpm-finsler,cjlw,kss,russell,rgt,schreck}
has primarily arisen
in the context of Lorentz-symmetry violation---a 
promising phenomenological signature 
in various theoretical approaches to quantum gravity~\cite{ksp,ncqed,spacetimevarying,qg,fn,bj,brane,modgrav}.

The Standard-Model Extension (SME) 
has been developed to describe Lorentz-violating effects 
at presently attainable energies 
regardless of their high-energy origin~\cite{sme1,sme2}. 
This framework has provided the basis 
for numerous experimental~\cite{DataTables,crays,photons,colliders,torsion,neutrinos} 
and theoretical~\cite{quant,renorm,fieldredef,OtherTheo} studies 
of Lorentz breaking.
In particular, 
one SME study has uncovered 
the incompatibility of Riemannian (and Riemann--Cartan) geometry 
with explicit Lorentz violation~\cite{sme2}, 
suggesting Riemann--Finsler spaces 
as the appropriate geometrical description of
Lorentz-violating physics in such situations.
Further theoretical evidence supporting this suggestion 
originates from the classical-particle limit of the SME~\cite{aknr}:
the motion of such particles
is effectively governed by geodesics in a Riemann--Finsler space~\cite{VAK_Finsler} 
despite the presumed underlying Riemannian structure.
This analysis 
also allows a partial classification of the emerging 
physically relevant Finsler structures~\cite{VAK_Finsler} 
according to the type of Lorentz breakdown.
For example, 
the SME's $a^{\mu}$ coefficient for fermions
leads to the familiar Randers space~\cite{gr}.

The present study is primarily concerned with Finsler $b$ space~\cite{VAK_Finsler},
which also emerges from the classical-particle limit of the SME.
To define $b$ space 
and other Finsler structures 
in a simplified form suitable for our present purposes, 
we note the following.
The aforementioned SME-inspired partial classification of Finsler spaces
is based on physical spacetime manifolds.
These possess indefinite metrics 
and are therefore pseudo-Riemannian.
For perturbative Lorentz violation, 
the indefinite-metric feature 
carries over to the Finsler-geometry interpretation, 
yielding pseudo-Finsler spaces.
In this work,
we are primarily concerned with 
the corresponding Finsler versions of these structures
with a positive-definite metric. 
In particular, 
the relevant base manifold for our present study is $\mathbb{R}^3$ 
endowed with the usual Euclidean metric. 
It then becomes unnecessary to distinguish between vectors and covectors.
Moreover, 
points on the manifold may be interpreted as vectors.
In general, 
we denote vectors as boldface letters.

With these consideration in mind, 
we recall the common notation for Finsler spaces involving
a scalar function $F$ on the tangent bundle:
integration of $F$ along a path on the manifold 
yields the Finsler path length
with geodesics defined as extremal paths 
between fixed points. 
For example, 
the Randers-space $F$ function is denoted by $F_a$ and given by
\beq
\label{Fa}
F_a(\mathbf{x},\mathbf{x}')=
\rho(\mathbf{x},\mathbf{x}')
+\alpha(\mathbf{x},\mathbf{x}')\,,
\eeq
where
\begin{align}
\label{rho_alpha}
\rho(\mathbf{x},\mathbf{x}') &=
\sqrt{\mathbf{x}'^2}\,,\nonumber\\
\alpha(\mathbf{x},\mathbf{x}') &=
\mathbf{a}(\mathbf{x})\cdot\mathbf{x}'\,.
\end{align}
Here, 
$\mathbf{x}=\mathbf{x}(\lambda)$ is a sufficiently well-behaved curve on the manifold,
$\mathbf{x}'=\mathbf{x}'(\lambda)$ is the corresponding velocity in the tangent space $T_{\mathbf{x}}M$, 
and $\mathbf{a}(\mathbf{x})$ is a prescribed (co)vector field.
The Finsler structure for $b$ space, 
on the other hand, 
is~\cite{VAK_Finsler}
\beq
\label{Fb}
F_b(\mathbf{x},\mathbf{x}')=
\rho(\mathbf{x},\mathbf{x}')
\pm\beta(\mathbf{x},\mathbf{x}')\,,
\eeq
where
\beq
\label{beta}
\beta(\mathbf{x},\mathbf{x}')=
\sqrt{\mathbf{b}(\mathbf{x})^2 \mathbf{x}'^2
-\big[\mathbf{b}(\mathbf{x})\cdot\mathbf{x}'\big]^2}\,,
\eeq
and $\mathbf{b}(\mathbf{x})$ is a given (co)vector field.
It has been argued that
$b$ space is in some sense complementary to Randers space~\cite{VAK_Finsler}.
In any case,
this Finsler $b$ structure emerges 
from the classical-particle limit 
of the SME's $b^\mu$ coefficient for fermions.
For completeness, 
we also recall the definition of the Finsler $ab$ structure~\cite{VAK_Finsler}
\beq
\label{Fab}
F_{ab}(\mathbf{x},\mathbf{x}')=
\rho(\mathbf{x},\mathbf{x}')
+\alpha(\mathbf{x},\mathbf{x}')
\pm\beta(\mathbf{x},\mathbf{x}')\,,
\eeq
which may be thought of as a combination of Randers and $b$ space.

In the present work, 
our primary focus is to find systems in classical physics that
are governed by the Finsler $b$ structure presented in Eq.~\rf{Fb}.
The identification of simple table-top systems 
described by the same set of equations 
as some original, less transparent system
has long been employed in physics research. 
Such analogues 
can give complementary perspectives on and further insight into the original system;
they may allow the application of ideas and methods from other fields of physics;
they may create opportunities for controlled experimentation within the analogue 
that can be translated back to the original system; 
and they can illustrate key physics ideas for educational purposes.
Examples of these types of analogues across various sub-disciplines of physics include
a number of Randers-space applications,
such as Zermelo navigation, optical metrics, and magnetic flow~\cite{RandersExamples};
modeling discrete spacetime-symmetry violations in meson oscillations
with mechanical and electric-circuit systems~\cite{discrete};
analogue gravity~\cite{AnalogueGravity};
models for PT-symmetric quantum mechanics~\cite{PTsymQM};
etc.

This Letter is structured as follows. 
Section~\ref{basics} reviews some basics about Finsler $b$ space 
and sets up our analysis.
In Sec.~\ref{bead}, 
we present a $b$-space application 
involving a bead sliding on a wire. 
A second system governed by the Finsler $b$ structure,
which is based on a transversely polarized magnetic chain,
is analyzed in Sec.~\ref{magnetic}.
Section~\ref{Geo} 
comments on some aspects of the corresponding geodesics.
A brief summary is contained in Sec.~\ref{sum}.


\section{Preliminary considerations}
\label{basics}

An important consideration in the construction of 
our classical-physics Finsler examples 
concerns the dimensionality of the manifold involved. 
For example,
the aforementioned Zermelo navigation problem 
can be formulated in two dimensions,
one of the key features that 
makes this Randers-space example 
particularly simple and intuitive.
In the present context, 
it would then seem natural to search also 
for simple, two-dimensional $b$-space examples.
However,  
the Finsler structures~\rf{Fa} and~\rf{Fb} 
are equivalent in two space dimensions~\cite{VAK_Finsler}.
Since our goal is to identify 
classical-physics analogues governed by 
true Finsler $b$ spaces,
nontrivial examples 
must involve at least three-dimensional manifolds.
Indeed, 
it can be shown~\cite{VAK_Finsler} that 
in three or more dimensions
the $b$-space Matsumoto torsion is non-vanishing.
Inequivalence to Randers space then follows directly from 
the Matsumoto--H\=oj\=o theorem~\cite{mh}.

Another consideration concerns 
the smoothness of $F_b$ 
because basic objects, 
such as the expression for the Finsler metric 
$g_{jk}=\frac{1}{2}\partial_{x'^j}\partial_{x'^k} F_b^2$,  
involve derivatives of $F_b$.
Inspection of $F_b$ reveals that
it is differentiable everywhere
with the exception of an extended slit $S$ 
given by $x'^j=\kappa\, b^j$, 
where $\kappa \in  \mathbb{R}$.
It has been conjectured that 
the geometry at this extended slit 
can be resolved 
using standard techniques 
for singularities of algebraic varieties~\cite{VAK_Finsler},
and progress towards such results 
has been made~\cite{dk14}.
From the perspective of the underlying SME quantum physics, 
this singularity can likely be addressed 
by introducing a spin variable~\cite{VAK_Finsler}.
In the present context of classical-physics analogues,
we focus on those solutions 
that do not involve the slit.

For $F_b$ to be a true Finsler structure, 
some additional conditions need to hold. 
These include homogeneity of degree one in $\mathbf{x}'$ 
and positive definiteness of both $F_b$ 
and the associated Finsler metric $g_{jk}$.
Homogeneity is established by inspection, 
and positive definiteness requires
\begin{equation}
\label{posdef}
|\mathbf{b}|<1
\end{equation}
for the lower sign in Eq.~(\ref{Fb}).
We note that 
this latter constraint
is compatible with the underlying motivation 
of the present analysis: 
$\mathbf{b}$ represents the analogue 
of the Lorentz-symmetry violating $b^\mu$ coefficient,
which must be perturbatively small 
on experimental grounds.

The construction of classical-physics analogues 
for Finsler geometry 
amounts to finding systems 
governed by a variational principle,
i.e., 
by the extremization of some integral.
At first sight, 
the Principle of Least Action 
applied to nonrelativistic mechanical systems 
would appear to be a natural starting point 
for the identification of such Finsler analogues. 
However, 
the presence of the mandatory kinetic-energy term 
in the Lagrangian $L$ 
seems to place tight restrictions 
on the overlap between $L$
and the set of possible Finsler structures $F$, 
and we have found that 
Lagrangians without a kinetic term 
will typically seem contrived. 

Another widely known physics application 
of the variational method
is Fermat's Principle in ray optics. 
It states that 
light rays between two fixed points $A$ and $B$ 
traverse the path of stationary optical length 
with respect to variations of the path, i.e.,
\begin{equation}
\label{Fermat}
\delta \int_A^B n(\mathbf{x},\mathbf{x}')\,ds=0\,.
\end{equation}
Here,
$s$ denotes the arc length 
of the light path $\mathbf{x}(s)$,
and $n(\mathbf{x},\mathbf{x}')$ is the refractive index,
which may depend on both the position $\mathbf{x}(s)$ 
and the propagation direction $\mathbf{x}'(s)$ 
of the light.
The prime expresses differentiation with respect to the argument.
The spatial dependence of 
the refractive index $n(\mathbf{x},\mathbf{x}')$ 
can now be selected to recover most Finsler structures.
Consider, 
for instance, 
the choice 
\begin{equation}
\label{n_choice}
n(\mathbf{x},\mathbf{x}')=1+\sqrt{\mathbf{b}^2-(\mathbf{b}\cdot\mathbf{e}_{\mathbf{x}'})^2}\;,
\end{equation}
where
$\mathbf{b}$ is a prescribed vector field, 
and $\mathbf{e}_{\mathbf{x}'}$ 
denotes the unit vector 
in the instantaneous propagation direction.
With this refractive index 
and a change of variables to a general,
well-behaved path parametrization $s=s(\lambda)$, 
Eq.~\rf{Fermat} 
takes the form of a classical system 
governed by Finsler $b$ space.
Although direction-dependent refractive indices 
are not uncommon in physics,
it would seem difficult 
to control experimentally the vector field $\mathbf{b}$: 
if only a constant $\mathbf{b}$ can be achieved, 
translational invariance would yield 
the perhaps less interesting case of 
straight-line light propagation (see Sec.~\ref{Geo}). 
In what follows,
we seek examples in which $\mathbf{b}$ 
is freely adjustable, 
at least in principle.

 
\section{Bead on a wire}
\label{bead}

As the first classical-physics analogue 
governed by a Finsler $b$-space geometry,
we consider a bead of mass $m$ that
slides in three space dimensions on a rough wire.
The bead's position vector as a function of time $t$ 
is denoted by $\mathbf{x}(t)$.
The bead fits tightly around the wire, 
and this tight fit by itself 
leads to a frictional force $\mathbf{F}_0$ 
that is constant in magnitude 
and directed against the beads velocity 
$\mathbf{v}(t)=\dot{\mathbf{x}}(t)$. 
The dot denotes the derivative with respect to time. 

The bead also experiences a prescribed external force
$\mathbf{F}_e(\mathbf{x})$. 
The physical origin of $\mathbf{F}_e(\mathbf{x})$ 
could, 
for example, 
be due to a gravitational field, 
an electric field (if the bead is charged), 
forces resulting from the flow of wind or water, etc. 
This external force 
and the bead's motion 
generally require a normal force $\mathbf{F}_n(\mathbf{x})$ 
perpendicular to $\dot{\mathbf{x}}(t)$ 
to keep the bead on the wire. 
This will lead to a further frictional force of magnitude
$\mu |\mathbf{F}_n(\mathbf{x})|$ opposing the bead's motion, 
where $\mu$ is a constant coefficient of kinetic friction.
This additional friction is assumed to be independent of $\mathbf{F}_0$.

The forces $\mathbf{F}_0$ and $\mathbf{F}_n(\mathbf{x})$, 
the friction coefficient $\mu$, 
and the initial conditions 
may not permit the bead to slide at all.
To circumvent this issue,
we allow for an additional force $\mathbf{F}_h$ that
can only act along the wire. 
Note in particular that 
this force can therefore not contribute to the friction.
We take $\mathbf{F}_h$ to be adjustable,
so that certain limiting cases,
such as quasistatic motion, 
can be achieved.
In a physical context,
$\mathbf{F}_h$ 
could for example provide the description of a hand 
moving the bead along the wire.

\begin{figure}
\centering
\includegraphics[width=0.95\columnwidth]{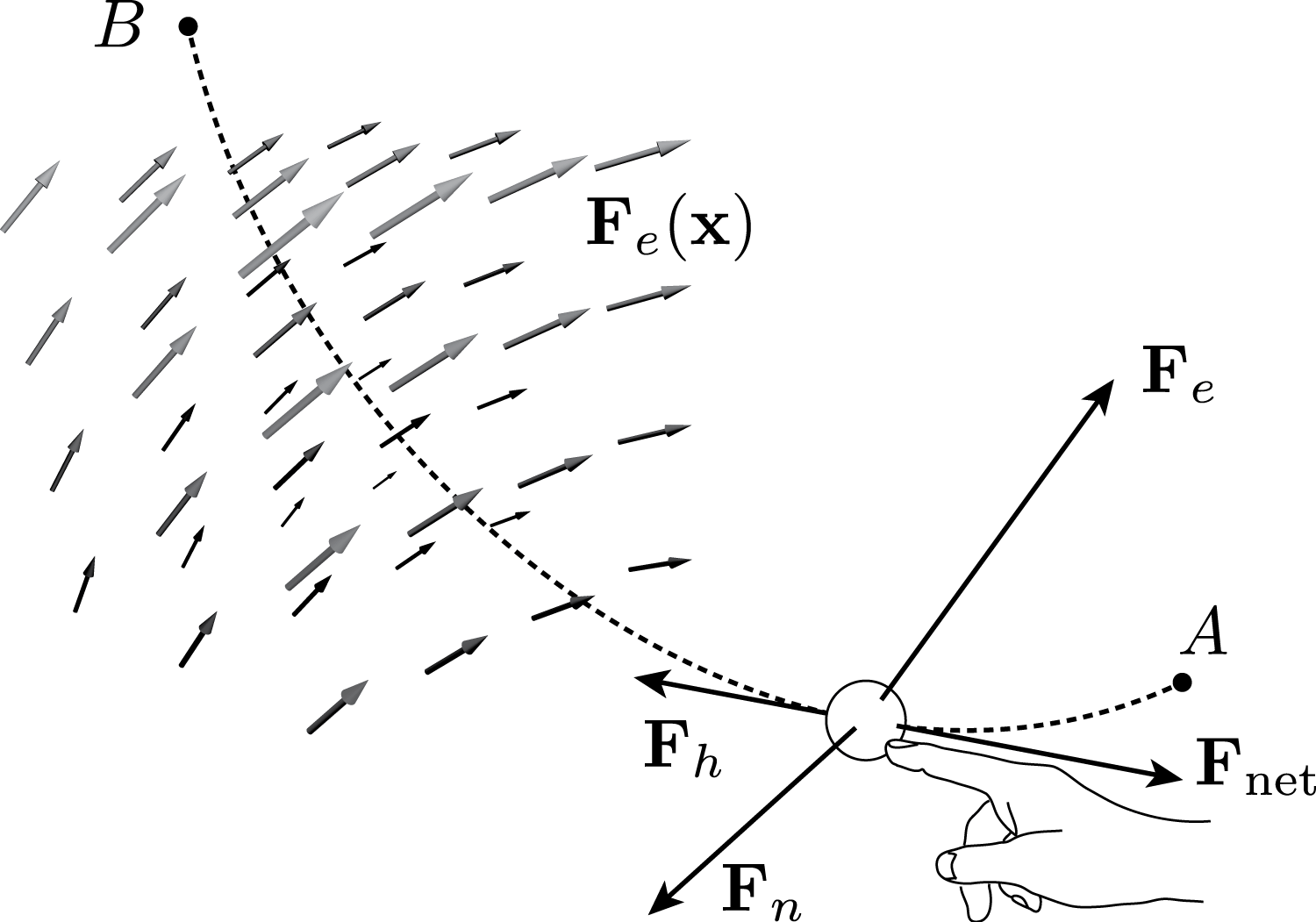}
\caption{Bead on a wire. 
The dashed line represents the wire 
with endpoints $A$ and $B$. 
The figure also shows 
the external force field $\mathbf{F}_e(\mathbf{x})$ 
and the free-body diagram for the bead. 
In the free-body diagram, 
the dependence of the forces on the position $\mathbf{x}$ 
is suppressed for brevity.}
\label{fig1}
\end{figure}

Given this classical-mechanics system,
we now consider the following situation.
Given two distinct points $A$ and $B$ in space that
represent the two ends of the wire,
the bead slides from $A$ to $B$.
Holding $A$ and $B$ fixed, 
how must the wire be bent 
for the energy loss $\Delta E$ due to friction
to be minimal? 
If the wire is not required to have a given constant length, 
$\Delta E$ can be expressed as 
\beq 
\label{DeltaE}
\Delta E = -\int_A^B\mathbf{F}_{\rm net}(\mathbf{x})\cdot d\mathbf{x}\,,
\eeq
where $\mathbf{F}_{\rm net}(\mathbf{x})$ denotes the net friction,
and the integration is understood to be carried out 
along a path $\mathbf{x}(t)$ 
with end points $A$ and $B$. 
The overall minus sign 
expresses our convention that 
the energy loss $\Delta E$ be positive.
Since the friction loss $\Delta E$ is associated with abrasive wear,
the problem of minimizing $\Delta E$ 
may, 
for instance, 
be motivated 
from an engineering standpoint 
that seeks to make a mechanical system last longer
by reducing abrasive wear. 
Our set-up is depicted schematically in Fig.~\ref{fig1}.

To find an explicit expression for $\mathbf{F}_{\rm net}(\mathbf{x})$ 
in terms of the given force field $\mathbf{F}_e(\mathbf{x})$,
it is useful to consider Newton's 2nd Law 
expressed in the Frenet--Serret frame that 
is comoving with the bead.
This frame is as usual composed of the unit vector $\mathbf{T}(t)$ 
tangent to the path $\mathbf{x}(t)$,
the unit normal vector $\mathbf{N}(t)$ 
pointing in the direction of $\dot{\mathbf{T}}(t)$ 
perpendicular to $\mathbf{T}(t)$, 
and the binormal vector 
$\mathbf{B}(t)\equiv\mathbf{T}(t)\times\mathbf{N}(t)$.
In the TNB basis,
the bead's acceleration 
is given by
$\ddot{\mathbf{x}}(t)=
\dot{v}(t)\mathbf{T}(t)+v(t)|\dot{\mathbf{T}}(t)|\mathbf{N}(t)$.
Suppressing the time and space dependence for brevity 
and denoting vector components in the $\mathbf{T}$, $\mathbf{N}$, and $\mathbf{B}$ directions with the respective superscripts $T$, $N$, and $B$,
Newton's 2nd Law takes the following form:
\bea
F_e^T+F_h^T-|\mathbf{F}_0|-\mu|\mathbf{F}_n| & = & m\dot{v}
\label{T}\\
F_e^N+F_n^N & = & m v |\dot{\mathbf{T}}|
\label{N}\\
F_e^B+F_n^B & = & 0\,.
\label{B}
\eea
Equations~\rf{N} and ~\rf{B} can be solved for 
the normal-force components $F_n^N$ and $F_n^B$.
With these components at hand,
the net frictional force becomes
\bea
\label{Fnet}
\mathbf{F}_{\rm net} & = &
-\big(|\mathbf{F}_0|+\mu|\mathbf{F}_n|\big)\mathbf{T}\nonumber\\
& = &-\bigg[|\mathbf{F}_0|+\mu\sqrt{\big(F_e^N-mv|\dot{\mathbf{T}}|\big)^2+\big(F_e^B\big)^2}\;\bigg]\;\mathbf{T}\,,\quad
\eea
where we have used that 
kinetic friction 
acts in the direction opposite to the velocity.

We are now in a position 
to express the integral~\rf{DeltaE} to be minimized 
in terms of $\mathbf{F}_e$.
The following change of variables 
$d\mathbf{x} = \dot{\mathbf{x}}\,dt=v\mathbf{T}\,dt$
gives
\beq 
\label{DeltaEexplicit}
\Delta E = \int_A^B \! v\; \bigg[|\mathbf{F}_0|+\mu\sqrt{\big(F_e^N-mv|\dot{\mathbf{T}}|\big)^2+\big(F_e^B\big)^2}\;\bigg]\; dt\,.
\eeq
We now proceed by 
considering two limiting cases of this expression, 
each governed by a Finsler $b$-space geometry.

The first case is the zero-mass limit, 
which eliminates the centripetal-force term in Eq.~\rf{DeltaEexplicit}. 
The square-root expression can then readily be identified with
the magnitude of the force
$\mathbf{F}_e^\perp=\mathbf{F}_e
-v^{-2}(\mathbf{F}_e\cdot\dot{\mathbf{x}})\,\dot{\mathbf{x}}$,
which is the component of $\mathbf{F}_e$ perpendicular to the wire, 
i.e., $\mathbf{F}_e^\perp\cdot\dot{\mathbf{x}}=0$. 
Recalling $v=\sqrt{\dot{\mathbf{x}}^2}$, 
we find
\beq 
\label{DeltaEmassless}
\Delta E = \int_A^B \bigg[|\mathbf{F}_0|\sqrt{\dot{\mathbf{x}}^2}
+\mu\sqrt{\mathbf{F}_e^2\,\dot{\mathbf{x}}^2
-\big(\mathbf{F}_e\cdot\dot{\mathbf{x}}\big)^2}\;\bigg]\; dt\,.
\eeq
for the energy loss due to friction.
It is apparent that 
with the identification $\mathbf{b}=\mu|\mathbf{F}_0|^{-1}\mathbf{F}_e$, 
we recover $F_b$
defined in Eq.~\rf{Fb} 
up to an overall factor. 
Thus, 
the zero-mass limit of our bead sliding on a wire 
indeed provides a classical-physics situation 
described by the Finsler $b$ structure 
when the energy loss is to be minimized.

Alternatively, 
we may consider 
the limit of quasistatic motion,
in which the bead is understood to move with a speed $v\to 0$.
This situation can, 
for example, 
be created 
if the bead
is moved slowly by hand along the wire, 
which can be modeled by
adjusting $\mathbf{F}_h$ 
such that the left-hand side of Eq.~\rf{T}, 
and thus $\dot{v}$, 
vanish. 
A slow initial speed then remains unchanged.
The direct implementation of the $v\to 0$ limit in Eq.~\rf{DeltaEexplicit}
lacks clarity 
because the integrand exhibits a multiplicative factor of $v$.
We therefore change integration variables 
from time $t$ 
to arc length $s$, 
so that the offending $v$ factor 
is absorbed into the integration measure $v\, dt = ds$.
The limit is now more transparent; 
it again eliminates the centripetal-force contribution:
\beq 
\label{DeltaEquasi}
\Delta E = \int_A^B \bigg[|\mathbf{F}_0|
+\mu\sqrt{\mathbf{F}_e^2
-\big(\mathbf{F}_e\cdot\mathbf{x}'\big)^2}\;\bigg]\; ds\,,
\eeq
where the prime denotes 
differentiation with respect to $s$.
With the same identification $\mathbf{b}=\mu|\mathbf{F}_0|^{-1}\mathbf{F}_e$ as before,
we recover up to an overall factor
the arc-length version of $F_b$ given in Eq.~\rf{Fb}. 
We therefore conclude that
the quasistatic limit of our bead-on-a-wire example
also provides a classical-physics analogue 
for Finsler $b$ space.


\section{Transversely magnetized chain}
\label{magnetic}

To establish a second classical-physics situation 
described by a Finsler $b$-space structure,  
we consider a catenary-type problem 
in three space dimensions.
More specifically, 
we examine the shape of a chain 
under the influence of magnetic forces 
in static equilibrium.

\begin{figure}
\centering
\includegraphics[width=0.95\columnwidth]{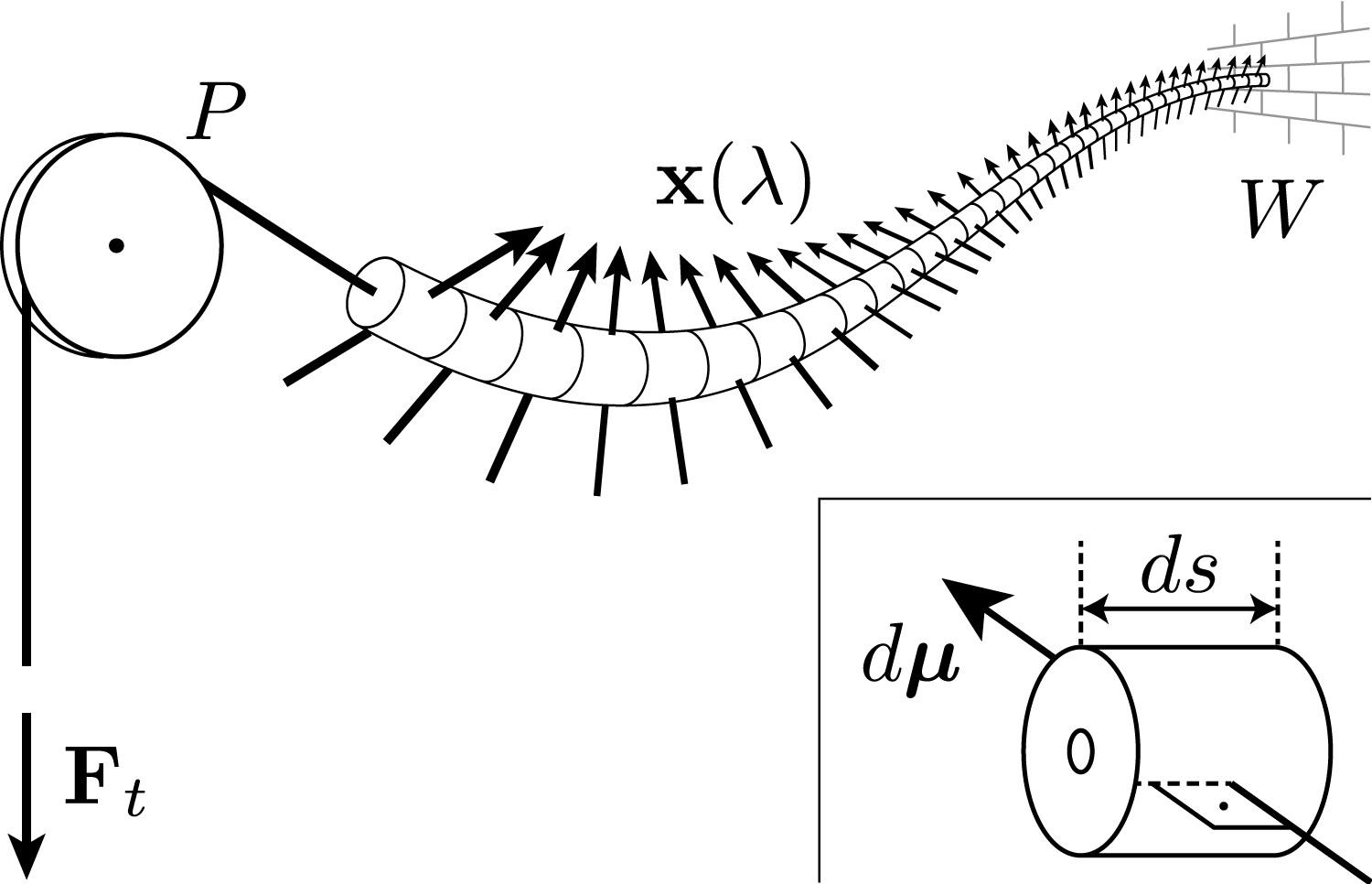}
\caption{Magnetic beads on a string. 
At the point $W$,
the string is attached to a wall.
Its other end hangs over a pulley at point $P$ 
and is held under a constant tension $\mathbf{F}_t$.
The beads are threaded tightly on the string.
They are free to rotate about the string
with their magnetic moments remaining perpendicular to the string.
The external $\mathbf{B}$ field is not shown.
The inset depicts the close-up of a single bead.
}
\label{fig2}
\end{figure}

The chain consists of beads 
that are threaded on a string.
This threading is tight in the sense that
neighboring beads are in contact with one another. 
This is to prevent motion of the beads along the string.
However, 
the beads are allowed to rotate freely and independently about the string.
The beads are identical. 
They carry a magnetic dipole moment $d\boldsymbol{\mu}$
that is oriented perpendicular to their symmetry axis 
and thus remains perpendicular to the string (see Fig.~\ref{fig2}).
At the point $W$,
this magnetic chain is held fixed, 
for example by attaching it to a wall.
The other end of the chain 
hangs over a pulley $P$ 
and is kept under constant tension $\mathbf{F}_t$. 
Such a force might,
for example, 
be provided by tensioning with a weight.  
This set-up is depicted in Fig.~\ref{fig2}.
The segment of the chain between $P$ and $W$ 
is exposed to 
a time-independent external magnetic field $\mathbf{B}(\mathbf{x})$. 
The question to be addressed 
concerns the equilibrium shape $\mathbf{x}(\lambda)$ 
of this magnetic chain. 

To find an equation for the equilibrium shape, 
we note that 
a static, stable configuration 
minimizes the total potential energy $E$ of the system. 
Our magnetic-chain set-up allows for various contributions to $E$. 
To make the problem tractable, 
we idealize the set-up by neglecting the following contributions.
First, 
we consider the string and the beads to be inelastic 
so as to preclude energy storage in the chain by stretching or compressing. 
As the second idealization,
we take the string and the beads to be massless
so that gravity can be ignored.
The third effect we neglect 
concerns the dipole--dipole interactions 
between different beads.\footnote{
The physical consistency of neglecting the dipole--dipole interactions 
may not be obvious.
Consider a set-up of discrete dipoles with finite $\Delta\mu$ 
separated by a finite $\Delta s$. 
This set-up closely resembles 
the actual physical situation 
at microscopic scales involving atoms.
Then, 
a single dipole in an external $\mathbf{B}$ field
possesses energy  $\Delta E_B\sim B\,\Delta\mu$, 
whereas its interaction with one of its neighboring dipoles contributes $\Delta E_{d-d}\sim\Delta\mu^2\,\Delta s^{-3}$. 
The $\mathbf{B}$ field can then always be selected 
such that $\Delta E_{d-d}$ is negligible and can be dropped, 
at least in principle. 
To obtain a mathematically more tractable continuum description, 
we can now approximate 
the fundamentally discrete sum over $\Delta E_B$ 
by a suitable integral.
Note, 
however,
that ignoring the atomistic nature of matter by 
taking $\Delta s \to 0$ 
would lead to a diverging $\Delta E_{d-d}$ 
for a constant linear dipole density 
$\Delta \mu/\Delta s$.}

In what follows, 
we focus on two contributions to the potential energy $E$.
One of these arises as a result of the tension $\mathbf{F}_t$ 
applied to the string.
This tension implies that 
any deformation of the chain
away from the straight-line configuration 
between $W$ and $P$
requires work to be done against $\mathbf{F}_t$, 
which is then stored as potential energy $E_t$ in the system.
Since $\mathbf{F}_t$ is constant, 
$E_t$ increases linearly with the length of chain 
drawn across the pulley.
Selecting a convenient energy zero, 
we thus find
\beq
\label{E_F}
E_t=F_t\int_W^P
\sqrt{\mathbf{x}'(\lambda)\cdot\mathbf{x}'(\lambda)}\;d\lambda\,,
\eeq
where $F_t$ denotes the magnitude of $\mathbf{F}_t$, 
and $\lambda$ is a general parameter along the chain 
such that $\mathbf{x}(\lambda)$
is sufficiently well-behaved.

The second contribution is a result of the magnetic interaction 
between the externally prescribed $\mathbf{B}(\mathbf{x})$
and the individual magnetic moments $d\boldsymbol{\mu}(\mathbf{x})$ 
of the beads. 
A single bead at $\mathbf{x}$ possesses
an energy $d E_B(\mathbf{x})$ given by~\cite{Jackson}:
\beq
\label{pot_En}
dE_B(\mathbf{x})=
-\mathbf{B}(\mathbf{x})\cdot d\boldsymbol{\mu}(\mathbf{x})\,.
\eeq
This expression shows that 
the net magnetic energy of the chain 
will consist of two contributions:
the orientation of the beads' magnetic momenta
and the shape of the chain.

As a first step, 
we consider an arbitrary fixed shape $\mathbf{x}(\lambda)$ 
of the chain.
This interaction~\rf{pot_En} 
causes a torque on the dipole 
in such a way 
as to align it with the magnetic field.\footnote{There can also be a net force on the dipole
if $\mathbf{B}$ is inhomogeneous.}
But as the beads are only allowed to rotate 
about one axis
(i.e., the fixed string),
the alignment is not necessarily perfect.  
Rather, 
the beads rotate such that 
the angle $\theta(\mathbf{x})$ 
between each bead's magnetic moment 
and the external magnetic field $\mathbf{B}(\mathbf{x})$ 
at the bead's position is minimized.
In particular, 
$0\leq\theta(\mathbf{x})\leq 90^\circ$ 
and $d\boldsymbol{\mu}(\mathbf{x})$ 
lies in the plane spanned by 
$\mathbf{B}(\mathbf{x})$ 
and the tangent $d\mathbf{x}$ 
to the curve $\mathbf{x}(\lambda)$.\footnote{On the slit described in Sec.~\ref{basics}, 
which corresponds to situations 
in which $\mathbf{B}(\mathbf{x})$ is tangential to $\mathbf{x}(\lambda)$,
this is ill-defined. 
As per our earlier discussion, 
we exclude such cases from our analysis.}
In this planar, stable configuration, 
we therefore have
\begin{equation}
\label{angle}
\cos \theta(\mathbf{x}) 
= \sin \phi(\mathbf{x})
=+\sqrt{1-\cos^2\phi(\mathbf{x})}\,,
\end{equation}
where $\phi(\mathbf{x})$ is the angle between 
$\mathbf{B}(\mathbf{x})$ and $d\mathbf{x}$.
Since the beads on our magnetic chain are identical, 
we can take the linear magnetic-moment density 
$\zeta=d\mu/ds$ to be constant
and write 
$d\mu(\mathbf{x})=\zeta \, ds$ 
for the magnitude $d\mu$ of the magnetic moment.
With these considerations, 
Eq.~\rf{pot_En} becomes
\begin{align}
\label{orientation}
dE_B(\mathbf{x})& = 
-B(\mathbf{x})\, \zeta \, ds \, \sqrt{1-\cos^2\phi(\mathbf{x})}\nonumber\\
& =  - \zeta \sqrt{
\mathbf{B}(\mathbf{x})^2\, d\mathbf{x}^2-\big[\mathbf{B}(\mathbf{x})\cdot d\mathbf{x}\big]^2}\,.
\end{align}
For a given shape $\mathbf{x}(\lambda)$, 
this yields
\begin{equation}
\label{magnetic_energy}
E_B = -\zeta \int_W^P
\sqrt{\mathbf{B}(\mathbf{x})^2\,\mathbf{x}'(\lambda)^2
-\big[\mathbf{B}(\mathbf{x})\cdot\mathbf{x}'(\lambda)\big]^2}\;d\lambda
\end{equation}
for the chain's net magnetic energy 
when the dipole beads are aligned in their stable 
orientation.
 
With the results~\rf{E_F} and~\rf{magnetic_energy} at hand,
the total energy $E$ of the chain is given by 
\begin{equation}
\label{total_E}
E=F_t\int_W^P\bigg[\sqrt{\mathbf{x}'^2}
-\frac{\zeta}{F_t}\sqrt{\mathbf{B}^2\mathbf{x}'^2
-(\mathbf{B}\cdot\mathbf{x}')^2}\bigg]\, d\lambda\,,
\end{equation}
where the dependence of $\mathbf{x}'$ 
and $\mathbf{B}$ on the curve $\mathbf{x}(\lambda)$ 
is understood. 
The equilibrium shape of our magnetic chain 
is now obtained by minimizing $E$. 
Note that with the identification 
$\mathbf{b}=\zeta\mathbf{B}/F_t$,
this system is governed by 
a Finsler $b$-space geometry.
We remark in passing that Eq.~\rf{total_E}
is subject to the physical requirement 
of $\nabla \cdot \mathbf{B}=0$ 
arising from the Maxwell equations.

The above catenary-type system 
can readily be modified or extended
to represent other Finsler structures. 
For example, 
a longitudinally magnetized chain 
would yield a Randers system.
A more elaborate bead structure 
involving transverse magnetic moments, 
as before, 
as well as longitudinal electric-dipole moments, 
would be governed by the $F_{ab}$ Finsler structure~\rf{Fab}
with external electric and magnetic fields 
playing the role of the prescribed 
$\mathbf{a}$ and $\mathbf{b}$ covector fields, 
respectively.


\section{Sample Geodesic Solution}
\label{Geo}

Thus far, 
we have identified two classical-physics systems that
are governed by Finsler $b$ space.
From a physics perspective, 
this merely corresponds to setting the problem up. 
Another interesting question concerns the solution
of this problem, 
i.e., 
finding the Finsler geodesics 
for a given $\mathbf{b}(\mathbf{x})$. 
This section briefly comments 
on this task.

Under mild assumptions, 
the $b$-space extremization problem
\begin{equation}
\label{extrem}
\delta\int_A^B 
F_b\big(\mathbf{x}(\lambda),\mathbf{x}'(\lambda)\big)
\,d\lambda=0
\end{equation}
can be converted into a system 
of three ordinary differential equations.
To this end,
we employ the usual Euler--Lagrange equations 
\begin{equation}
\label{EL_Eqs}
\frac{\partial F_b}{\partial x_j}=
\frac{d}{d\lambda}\frac{\partial F_b}{\partial x_j'}
\end{equation}
to obtain:
\begin{equation}
\label{EoM}
\nabla \beta = 
\pm\Bigg(\frac{\mathbf{x}''}{\rho}
-\frac{\mathbf{x}'\cdot\mathbf{x}''}{\rho^3}\,\mathbf{x}'\Bigg)
+\frac{d}{d\lambda}\,\frac{1}{\beta}
\big[b^2\mathbf{x}'-(\mathbf{b}\cdot\mathbf{x}')\,\mathbf{b}\big]\,.
\end{equation}
Here, 
the upper and lower sign choice
corresponds to that in Eq.~(\ref{Fb}).
In general, 
this system of equations is likely to be intractable. 
However, 
we may find solutions 
for special choices for $\mathbf{b}(\mathbf{x})$.

Consider first the case of a constant $\mathbf{b}$.
This case can be used to illustrate that
our simple analogue models 
can provide insight into Finsler $b$ space. 
In our magnetized-chain example, 
$\mathbf{b}=\textrm{const.}$ 
corresponds to a homogeneous magnetic field,
which does not exert any forces on the dipoles.
We therefore expect that 
the equilibrium shape of the chain 
is a straight line.

To see this more rigorously
note that 
the left-hand side of Eq.~\rf{EL_Eqs} vanishes, 
and thus
\begin{equation}
\label{constb1}
\frac{\partial F_b}{\partial \mathbf{x}'}=
\Bigg(\frac{1}{\rho}\pm\frac{b^2}{\beta}\Bigg)\,\mathbf{x}'\mp\frac{\mathbf{x}'\cdot\mathbf{b}}{\beta}\,\mathbf{b}
=\mathbf{C}\,,
\end{equation}
where $\mathbf{C}$ is a constant. 
Contraction of this equation with $\mathbf{b}$
yields
\begin{equation}
\label{constb2}
\mathbf{x}'\cdot\mathbf{b}=
\rho\, \mathbf{C}\cdot\mathbf{b}=\rho\, c\,,
\end{equation}
with the constant $c$ defined as 
$c\equiv\mathbf{C}\cdot\mathbf{b}$. 
Using Eq.~\rf{constb2}, 
all occurrences of the scalar product $\mathbf{x}'\cdot\mathbf{b}$ 
in Eq.~\rf{constb1} 
can be traded for $\rho c$. 
A rearrangement of terms in Eq.~\rf{constb1} then gives
\begin{equation}
\label{constb3}
\frac{\mathbf{x}'}{\rho}=
\Bigg(1\pm\frac{b^2}{\sqrt{b^2-c^2}}\Bigg)^{-1}
\Bigg(\mathbf{C}\pm\frac{c}{\sqrt{b^2-c^2}}\,\mathbf{b}\Bigg)\,.
\end{equation}
Since the right-hand side of this equation is constant,
we conclude that the unit vector in tangential direction 
$\mathbf{x}'/\rho$ is constant as well, 
so $\mathbf{x}(\lambda)$
must be a straight line 
with respect to our Euclidean base manifold. 
This result is consistent with 
the flat-spacetime SME's pseudo-Finsler case, 
in which free fermions 
with a constant Lorentz-violating $b^\mu$ coefficient 
also travel along straight paths.

Although straight-line solutions are analogous to
fermion propagation in the Minkowski-space SME,
it might also be interesting 
to find sample solutions that 
do not represent a straight path 
in the Euclidean base manifold. 
To this end, 
consider a $\mathbf{b}(\mathbf{x})$ field with the following 
cylinder symmetry:
\begin{equation}
\label{cylb}
\mathbf{b}(\mathbf{x})=
\big[b_0 \exp(-\sigma \varrho) - 1 \big]\,\hat{\boldsymbol{\varrho}}\,,
\end{equation}
where $\varrho$ denotes the radial cylinder coordinate 
and $\hat{\boldsymbol{\varrho}}$ is the unit vector associated to $\varrho$.  
The range for the parameters $b_0$ and $\sigma$ 
is given by
\begin{equation}
\label{bsigmaconstraints1}
1 < b_0 < 2\,,\qquad
0 < \sigma \,.
\end{equation}
These constraints guarantee 
the positivity requirement~(\ref{posdef}) 
and the existence of a nonvanishing cylindrical region
\begin{equation}
\label{region1}
0<\varrho<\sigma^{-1}\ln b_0\,,
\end{equation}
in which 
$\mathbf{b}(\mathbf{x})$ 
points {\it away} from the $z$ axis. 
It is this cylindrical region, 
in which we seek helix solutions of the general form
\begin{equation}
\label{helix}
\mathbf{x}(\lambda)=
R\cos\lambda\;\hat{\mathbf{x}}
+R\sin\lambda\;\hat{\mathbf{y}}
+\frac{h}{2\pi}(\lambda-\lambda_0)\;\hat{\mathbf{z}}\,.
\end{equation}
To be located in the region~(\ref{region1}),
the helix radius $R>0$ must obey 
\begin{equation}
\label{R1}
0<R<\sigma^{-1}\ln b_0\,<\sigma^{-1}.
\end{equation}
Here, 
the last of these inequalities 
follows from the requirement~(\ref{bsigmaconstraints1}).
The parameter $h\in\mathbb{R}$ 
controls the handedness and the pitch of the helix, 
and $\lambda_0\in\mathbb{R}$ the offset along the $z$ axis.

With these definitions, 
the geodesic equation~(\ref{EoM}) 
with the upper, positive sign selected 
is satisfied if
\begin{equation}
\label{h}
|h|=2\pi R\sqrt{(\sigma R)^{-1} -1}\,.
\end{equation} 
Together with the relation~(\ref{R1}),
this implies 
\begin{equation}
\label{hrange}
|h|>2\pi\sigma^{-1}\ln b_0\,.
\end{equation}
In view of the parameter ranges~(\ref{bsigmaconstraints1}), it then follows that
this solution is indeed a proper helix,
and not a circle $h=0$.
It is also apparent that 
both handedness choices are acceptable.
Note that $\lambda_0$ is left unconstrained 
by the geodesic equation,
as expected 
from translational symmetry of $\mathbf{b}(\mathbf{x})$
along the $z$ axis.
We also note that 
the Finsler $b$ structure
is left invariant under 
$\mathbf{b}(\mathbf{x})\to-\mathbf{b}(\mathbf{x})$. 
This observation implies that 
the established helix solution 
also solves the upper-sign version of Eq.~(\ref{EoM}) 
for an external field 
$\mathbf{b}(\mathbf{x})=
-\big[b_0 \exp(-\sigma \varrho) - 1 \big]\,\hat{\boldsymbol{\varrho}}$.

The derivation of the above solution 
has assumed the case 
in which $\beta$ enters the Finler structure~(\ref{Fb}) 
with a plus sign. 
To find a nontrivial solution for the other sign choice, 
we consider the same expression~(\ref{cylb}) for the external field, 
but we focus on the region in which $\mathbf{b}(\mathbf{x})$
points {\it towards} the $z$ axis.
Positivity~(\ref{posdef}) holds if
\begin{equation}
\label{bsigmaconstraints}
0 < b_0 < 2\,,\qquad
0 < \sigma \,.
\end{equation}
As opposed to the analogous expressions~(\ref{bsigmaconstraints1}), 
this range has been chosen solely to guarantee the positivity condition~(\ref{posdef});
no additional restrictions are needed 
since there is always a nonvanishing region 
\begin{equation}
\label{region}
\textrm{max}(0,\sigma^{-1}\ln b_0)<\varrho\,,
\end{equation}
with the desired directionality of $\mathbf{b}(\mathbf{x})$.
We again seek helix solutions of the form~(\ref{helix}),
so we must have
\begin{equation}
\label{R2}
\textrm{max}(0,\sigma^{-1}\ln b_0)<R
\end{equation}
to be in the region of interest. 
The geodesic equation, 
now with the lower, negative sign, 
yields again Eq.~(\ref{h}).
However, 
since the parameter space for $R$ 
has changed from the range~(\ref{R1}) 
to the range~(\ref{R2}), 
a real-valued $|h|$  is no longer automatic;
we must impose
\begin{equation}
\label{Rbound}
R\leq\sigma^{-1}\,.
\end{equation}
Nevertheless, 
the inequalities~(\ref{R2}) and~(\ref{Rbound}) 
admit a finite range for $R$.
They now also allow the case $h=0$.
As before, 
$\lambda_0$ is left undetermined, 
and invariance under 
$\mathbf{b}(\mathbf{x})\to-\mathbf{b}(\mathbf{x})$
implies that the flipped external field 
$\mathbf{b}(\mathbf{x})=
-\big[b_0 \exp(-\sigma \varrho) - 1 \big]\,\hat{\boldsymbol{\varrho}}$
leads to the same helix solution.


\section{Summary}
\label{sum} 

This work has considered  
a particular class of Finsler spaces 
called $b$ space.
This Finsler class 
is complementary to the widely known Randers space 
and arises 
in its pseudo-Finsler version 
as the classical limit 
of SME fermion propagation. 
We have shown that 
three-dimensional $b$ spaces 
also have other applications 
in conventional classical physics.

One of these applications 
involves a bead 
sliding under friction and external forces on a wire.
The shape of the wire that
minimizes abrasive forces 
follows a Finsler $b$-space geodesic.
In this example,
the $\beta$ term enters the expression~(\ref{Fb}) for $F_b$
with a plus sign. 
The physical origin of this sign 
arises from the fact that 
this $\beta$ term describes heat 
generated by friction, 
which is always dissipative.

The second example 
we have discussed 
concerns a transversely magnetized chain 
in an external magnetic field: 
the static-equilibrium configuration of the chain 
is governed by a length-minimizing path 
with respect to a Finslerian $b$ metric. 
In this set-up,
$\beta$ enters the expression~(\ref{Fb}) for $F_b$
with a negative sign 
for the following physical reason. 
The $\beta$ term 
models the magnetic potential energy of the beads 
according to Eq.~(\ref{pot_En}).
Since the dipoles are free to rotate about the string,
the lowest-energy configuration 
(away from the slit)
is always characterized by 
an acute angle between the dipole and the $\mathbf{B}$ field.
This implies a negative magnetic energy for each bead.

The general solution for the geodesic curve 
for a given $\mathbf{b}(\mathbf{x})$ 
would be interesting, 
but appears to be difficult to determine.
However, 
for the sample cases of a constant $\mathbf{b}$ 
and a $\mathbf{b}(\mathbf{x})$ 
with a particular cylinder symmetry 
we have identified nontrivial geodesics.


\section*{Acknowledgements}

We thank the unknown referee
for various improvements to our analysis.
This work was supported in part
by Indiana University's STARS program,
by the National Science Foundation
under the REU program,
and by the Indiana University Center for Spacetime Symmetries
under an IUCRG grant.



\section*{References}

\begin{thebibliography}{xx}

\bibitem{br}
B.~Riemann,
{\it \"Uber die Hypothesen welche der Geometrie zu Grunde liegen},
in R.~Baker, C.~Christensen, and H.~Orde,
{\it Bernhard Riemann, Collected Papers},
Kendrick Press, Heber City, Utah, 2004.

\bibitem{pf}
P.~Finsler,
{\it \"Uber Kurven und Fl\"achen in allgemeinen R\"aumen},
University of G\"ottingen dissertation, 1918;
Verlag Birkh\"auser, Basel, Switzerland, 1951.

\bibitem{bcs}
D.~Bao, S.-S.~Chern, and Z.~Shen,
{\it An Introduction to Riemann--Finsler Geometry},
Springer, New York, 2000.

\bibitem{sczs}
S.-S.~Chern and Z.~Shen,
{\it Riemann-Finsler Geometry},
World Scientific, Singapore, 2005.

\bibitem{gyb}
G.Yu.~Bogoslovsky and H.F.~Goenner,
Gen.\ Rel.\ Grav.\  {\bf 31}, 1383 (1999);
Phys.\ Lett.\ A {\bf 244}, 222 (1998);
Gen.\ Rel.\ Grav.\  {\bf 31}, 1565 (1999);
Phys.\ Lett.\ A {\bf 323}, 40 (2004);
G.Yu.~Bogoslovsky,
Phys.\ Lett.\ A {\bf 350}, 5 (2006);
Int.\ J.\ Geom.\ Meth.\ Mod.\ Phys.\ {\bf 9}, 1250007 (2012);
V.~Balan {\it et al.}, 
J.\ Mod.\ Phys.\  {\bf 3}, 1314 (2012).

\bibitem{Girelli07} 
F.~Girelli, S.~Liberati, and L.~Sindoni,
Phys.\ Rev.\ D {\bf 75}, 064015 (2007);
L.~Sindoni,
Phys.\ Rev.\ D {\bf 77}, 124009 (2008).

\bibitem{Gibbons07} 
G.W.~Gibbons, J.~Gomis, and C.N.~Pope,
Phys.\ Rev.\ D {\bf 76}, 081701 (2007).

\bibitem{lph}
C.~L\"ammerzahl, D.~Lorek, and H.~Dittus,
Gen.\ Rel.\ Grav.\  {\bf 41}, 1345 (2009);
C.~L\"ammerzahl, V.~Perlick, and W.~Hasse,
Phys.\ Rev.\ D {\bf 86}, 104042 (2012).

\bibitem{jsmv}
J.~Sk\'akala and M.~Visser,
Int.\ J.\ Mod.\ Phys.\ D {\bf 19}, 1119 (2010);
J.\ Geom.\ Phys.\  {\bf 61}, 1396 (2011);
Class.\ Quant.\ Grav.\  {\bf 28}, 065007 (2011).

\bibitem{VAK_Finsler} 
V.A.~Kosteleck\'y,
Phys.\ Lett.\ B {\bf 701}, 137 (2011).

\bibitem{sv}
S.I.~Vacaru,
Class.\ Quant.\ Grav.\ {\bf 28}, 215001 (2011).

\bibitem{V11} 
N.~Voicu,
PIER {\bf 113}, 83 (2011).

\bibitem{rsv}
J.M.~Romero, O.~Sanchez-Santos, and J.D.~Vergara,
Phys.\ Lett.\ A {\bf 375}, 3817 (2011).

\bibitem{pw11}
C.~Pfeifer and M.N.R.~Wohlfarth,
Phys.\ Rev.\ D {\bf 84}, 044039 (2011).

\bibitem{js}
E.~Caponio, M.A.~Javaloyes, and A.~Masiello,
Math.\ Ann.\  {\bf 351}, 365 (2011);
E.~Caponio, M.A.~Javaloyes, and M.~S\'anchez,
Rev.\ Mat.\ Iberoam.\  {\bf 27}, 919 (2011);
M.A.~Javaloyes and M.~S\'anchez,
arXiv:1111.5066,
to appear in Ann.\ Sc.\ Norm.\ Sup.\ Pisa, 
DOI Number: 10.2422/2036-2145.201203\_002;
A.B.~Aazami and M.A.~Javaloyes,
arXiv:1410.7595;
M.A.~Javaloyes and H.~Vit\'orio,
arXiv:1412.0465.

\bibitem{nm}
N.E.~Mavromatos,
Phys.\ Rev.\ D {\bf 83}, 025018 (2011);
N.E.~Mavromatos, S.~Sarkar, and A.~Vergou,
Phys.\ Lett.\ B {\bf 696}, 300 (2011);
N.E.~Mavromatos, V.A.~Mitsou, S.~Sarkar, and A.~Vergou,
Eur.\ Phys.\ J.\ C {\bf 72}, 1956 (2012).

\bibitem{dcpm-finsler}
D.~Colladay and P.~McDonald, 
Phys.\ Rev.\ D {\bf 85}, 044042 (2012).

\bibitem{cjlw}
Z.~Chang, X.~Li, and S.~Wang,
arXiv:1201.1368;
Z.~Chang and S.~Wang,
Eur.\ Phys.\ J.\ C {\bf 72}, 2165 (2012).

\bibitem{kss}
P.~Stavrinos,
Gen.\ Rel.\ Grav.\  {\bf 44}, 3029 (2012);
A.P.~Kouretsis, M.~Stathakopoulos, and P.C.~Stavrinos,
Phys.\ Rev.\ D {\bf 86}, 124025 (2012).

\bibitem{russell}
V.A.~Kosteleck\'y, N.~Russell, and R.~Tso,
Phys.\ Lett.\ B {\bf 716}, 470 (2012);
N.~Russell,
arXiv:1501.02490 [hep-th].

\bibitem{rgt}
R.G.~Torrom\'e,
arXiv:1207.3791.

\bibitem{Silva:2013qxa} 
J.E.G.~Silva and C.A.S.~Almeida,
arXiv:1309.4671 [hep-th].

\bibitem{schreck}
M.~Schreck,
arXiv:1405.5518 [hep-th];
arXiv:1409.1539 [hep-th].

\bibitem{ksp}
V.A.~Kosteleck\'y and S.~Samuel,
Phys.\ Rev.\ D {\bf 39}, 683 (1989);
V.A.~Kosteleck\'y and R.~Potting,
Nucl.\ Phys.\ B {\bf 359}, 545 (1991);
K.~Hashimoto and M.~Murata,
PTEP {\bf 2013}, 043B01 (2013).

\bibitem{ncqed}
I.~Mocioiu, M.~Pospelov, and R.~Roiban,
Phys.\ Lett.\ B {\bf 489}, 390 (2000);
S.M.~Carroll {\it et al.},
Phys.\ Rev.\ Lett.\  {\bf 87}, 141601 (2001);
C.E.~Carlson, C.D.~Carone, and R.F.~Lebed,
Phys.\ Lett.\ B {\bf 518}, 201 (2001);
A.~Anisimov, T.~Banks, M.~Dine, and M.~Graesser,
Phys.\ Rev.\ D {\bf 65}, 085032 (2002).

\bibitem{spacetimevarying}
V.A.~Kostelecky, R.~Lehnert, and M.J.~Perry,
Phys.\ Rev.\ D {\bf 68}, 123511 (2003);
R.~Jackiw and S.-Y.~Pi,
Phys.\ Rev.\ D {\bf 68}, 104012 (2003);
O.~Bertolami, R.~Lehnert, R.~Potting, and A.~Ribeiro,
Phys.\ Rev.\ D {\bf 69}, 083513 (2004).

\bibitem{qg}
J.~Alfaro, H.A.~Morales-T\'ecotl, and L.F.~Urrutia,
Phys.\ Rev.\ Lett.\  {\bf 84}, 2318 (2000);
F.R.~Klinkhamer and C.~Rupp,
Phys.\ Rev.\ D {\bf 70}, 045020 (2004);
G.~Amelino-Camelia {\it et al.},
AIP Conf.\ Proc.\  {\bf 758}, 30 (2005);
N.E.~Mavromatos,
Lect.\ Notes Phys.\  {\bf 669}, 245 (2005).

\bibitem{fn}
C.D.~Froggatt and H.B.~Nielsen,
hep-ph/0211106.

\bibitem{bj}
J.D.~Bjorken,
Phys.\ Rev.\ D {\bf 67}, 043508 (2003).

\bibitem{brane}
C.P.~Burgess {\it et al.}, 
JHEP {\bf 0203}, 043 (2002);
A.R.~Frey,
JHEP {\bf 0304}, 012 (2003);
J.M.~Cline and L.~Valc\'arcel,
JHEP {\bf 0403}, 032 (2004).

\bibitem{modgrav}
V.A.~Kosteleck\'y and S.~Samuel,
Phys.\ Rev.\ D {\bf 42}, 1289 (1990);
N.~Arkani-Hamed {\it et al.},
JHEP {\bf 0405}, 074 (2004);
M.V.~Libanov and V.A.~Rubakov,
JHEP {\bf 0508}, 001 (2005);
G.~Dvali, O.~Pujolas, and M.~Redi,
Phys.\ Rev.\ D {\bf 76}, 044028 (2007);
S.~Dubovsky, P.~Tinyakov, and M.~Zaldarriaga,
JHEP {\bf 0711}, 083 (2007).

\bibitem{sme1}
D.~Colladay and V.A.~Kosteleck\'y,
Phys.\ Rev.\ D {\bf 55}, 6760 (1997);
Phys.\ Rev.\ D {\bf 58}, 116002 (1998).

\bibitem{sme2} 
V.A.~Kosteleck\'y,
Phys.\ Rev.\ D {\bf 69}, 105009 (2004).

\bibitem{DataTables}
For an overview of various tests of Lorentz symmetry see
V.A.~Kosteleck\'y and N.~Russell,
Rev.\ Mod.\ Phys.\  {\bf 83}, 11 (2011)
[2014 edition: arXiv:0801.0287v7].

\bibitem{crays}
See, e.g.,
S.R.~Coleman and S.L.~Glashow,
Phys.\ Lett.\ B {\bf 405}, 249 (1997);
T.~Jacobson, S.~Liberati, and D.~Mattingly,
Phys.\ Rev.\ D {\bf 66}, 081302 (2002);
R.~Lehnert,
Phys.\ Rev.\ D {\bf 68}, 085003 (2003);
F.R.~Klinkhamer and M.~Schreck,
Phys.\ Rev.\ D {\bf 78}, 085026 (2008);
S.T.~Scully and F.W.~Stecker,
Astropart.\ Phys.\  {\bf 31}, 220 (2009).

\bibitem{photons}
V.A.~Kosteleck\'y and M.~Mewes,
Phys.\ Rev.\ D {\bf 66}, 056005 (2002);
R.~Lehnert and R.~Potting,
Phys.\ Rev.\ Lett.\  {\bf 93}, 110402 (2004);
Phys.\ Rev.\ D {\bf 70}, 125010 (2004);
V.A.~Kosteleck\'y and M.~Mewes,
Phys.\ Rev.\ D {\bf 80}, 015020 (2009);
A.A.~Abdo {\it et al.}  
Science {\bf 323}, 1688 (2009).

\bibitem{colliders}
V.A.~Kosteleck\'y and R.~Potting,
Phys.\ Rev.\ D {\bf 51}, 3923 (1995);
M.A.~Hohensee {\it et al.},
Phys.\ Rev.\ Lett.\  {\bf 102}, 170402 (2009);
Phys.\ Rev.\ D {\bf 80}, 036010 (2009);
B.~Altschul,
Phys.\ Rev.\  D {\bf 80}, 091901 (2009);
G.~Amelino-Camelia {\it et al.},
Eur.\ Phys.\ J.\ C {\bf 68}, 619 (2010);
J.-P.~Bocquet {\it et al.},
Phys.\ Rev.\ Lett.\  {\bf 104}, 241601 (2010);
A.~Di Domenico {\it et al.},
Found.\ Phys.\  {\bf 40}, 852 (2010).

\bibitem{torsion}
V.A.~Kosteleck\'y, N.~Russell, and J.~Tasson,
Phys.\ Rev.\ Lett.\  {\bf 100}, 111102 (2008);
R.~Lehnert, W.M.~Snow, and H.~Yan,
Phys.\ Lett.\ B {\bf 730}, 353 (2014).

\bibitem{neutrinos}
V.~Barger, D.~Marfatia, and K.~Whisnant,
Phys.\ Lett.\ B {\bf 653}, 267 (2007);
P.~Adamson {\it et al.},
Phys.\ Rev.\ Lett.\  {\bf 105}, 151601 (2010);
A.G.~Cohen and S.L.~Glashow,
Phys.\ Rev.\ Lett.\  {\bf 107}, 181803 (2011);
V.A.~Kosteleck\'y and M.~Mewes,
Phys.\ Rev.\ D {\bf 85}, 096005 (2012);
T.~Katori,
Mod.\ Phys.\ Lett.\ A {\bf 27}, 1230024 (2012);
J.S.~D\'iaz, V.A.~Kosteleck\'y, and R.~Lehnert,
Phys.\ Rev.\ D {\bf 88}, 071902 (2013).

\bibitem{quant}
V.A.~Kosteleck\'y and R.~Lehnert,
Phys.\ Rev.\ D {\bf 63}, 065008 (2001);
C.~Adam and F.R.~Klinkhamer,
Nucl.\ Phys.\ B {\bf 657}, 214 (2003);
C.M.~Reyes, L.F.~Urrutia, and J.D.~Vergara,
Phys.\ Rev.\ D {\bf 78}, 125011 (2008);
M.A.~Hohensee, D.F.~Phillips, and R.L.~Walsworth,
arXiv:1210.2683 [quant-ph];
D.~Colladay, P.~McDonald, and R.~Potting,
Phys.\ Rev.\ D {\bf 89}, no. 8, 085014 (2014);
C.~Hernaski,
Phys.\ Rev.\ D {\bf 90}, no. 12, 124036 (2014).

\bibitem{renorm}
V.A.~Kosteleck\'y, C.D.~Lane, and A.G.M.~Pickering,
Phys.\ Rev.\  D {\bf 65}, 056006 (2002);
G.~de Berredo-Peixoto and I.L.~Shapiro,
Phys.\ Lett.\ B {\bf 642}, 153 (2006);
D.~Colladay and P.~McDonald,
Phys.\ Rev.\ D {\bf 79}, 125019 (2009);
Phys.\ Rev.\ D {\bf 77}, 085006 (2008);
Phys.\ Rev.\ D {\bf 75}, 105002 (2007);
A.~Ferrero and B.~Altschul,
Phys.\ Rev.\ D {\bf 84}, 065030 (2011).

\bibitem{fieldredef}
D.~Colladay and P.~McDonald,
J.\ Math.\ Phys.\  {\bf 43}, 3554 (2002);
R.~Lehnert,
Phys.\ Rev.\ D {\bf 74}, 125001 (2006);
Rev.\ Mex.\ Fis.\  {\bf 56}, 469 (2010);
B.~Altschul,
J.\ Phys.\ A {\bf 39}, 13757 (2006).

\bibitem{OtherTheo}
R.~Jackiw and V.A.~Kosteleck\'y,
Phys.\ Rev.\ Lett.\  {\bf 82}, 3572 (1999);
M.~P\'erez-Victoria,
Phys.\ Rev.\ Lett.\  {\bf 83}, 2518 (1999);
R.~Lehnert,
J.\ Math.\ Phys.\  {\bf 45}, 3399 (2004);
B.~Altschul,
Phys.\ Rev.\ D {\bf 73}, 036005 (2006);
Q.G.~Bailey and V.A.~Kosteleck\'y,
Phys.\ Rev.\ D {\bf 74}, 045001 (2006);
A.J.~Hariton and R.~Lehnert,
Phys.\ Lett.\ A {\bf 367}, 11 (2007);
V.A.~Kosteleck\'y and J.D.~Tasson,
Phys.\ Rev.\ D {\bf 83}, 016013 (2011);
R.~Potting,
Phys.\ Rev.\ D {\bf 85}, 045033 (2012);
M.~Cambiaso, R.~Lehnert, and R.~Potting,
Phys.\ Rev.\ D {\bf 85}, 085023 (2012);
Phys.\ Rev.\ D {\bf 90}, no. 6, 065003 (2014);
M.~Schreck,
Phys.\ Rev.\ D {\bf 89}, no. 10, 105019 (2014);
Phys.\ Rev.\ D {\bf 90}, no. 8, 085025 (2014).

\bibitem{aknr}
V.A.~Kosteleck\'y and N.~Russell,
Phys.\ Lett.\ B {\bf 693}, 443 (2010).

\bibitem{gr}
G.~Randers,
Phys.\ Rev.\ {\bf 59}, 195 (1941).

\bibitem{RandersExamples}
T.~Sunada, 
Proc.\ KAIST Math.\ Workshop {\bf 8}, 93 (1993);
Z.~Shen,
Canad.\ J.\ Math.\ {\bf 55}, 112 (2003);
D.~Bao, C.~Robles, and Z.~Shen,
J.\ Diff.\ Geom.\ {\bf 66}, 377 (2004);
D.~Bao and C.~Robles,
in D.~Bao, R.L.~Bryant, S.-S.~Chern, and Z.~Shen, eds.,
{\it A Sampler of Riemann-Finsler Geometry},
Cambridge University Press, Cambridge, 2004;
G.W.~Gibbons, C.A.R.~Herdeiro, C.M.~Warnick, and M.C.~Werner,
Phys.\ Rev.\ D {\bf 79}, 044022 (2009).

\bibitem{discrete} 
B.~Winstein,
in K.~Winter, ed., 
{\it Festi-Val---Festschrift for Val Telegdi} 
Elsevier, Amsterdam, 1988;
J.L.~Rosner,
Am.\ J.\ Phys.\  {\bf 64}, 982 (1996);
J.L.~Rosner and S.A.~Slezak,
Am.\ J.\ Phys.\  {\bf 69}, 44 (2001);
V.A.~Kosteleck\'y and A.~Roberts,
Phys.\ Rev.\ D {\bf 63}, 096002 (2001);
M.~Caruso, H.~Fanchiotti, and C.A.~Garcia Canal,
Annals Phys.\  {\bf 326}, 2717 (2011);
K.R.~Schubert and J.~Stiewe,
J.\ Phys.\ G {\bf 39}, 033101 (2012);
A.~Reiser, K.R.~Schubert, and J.~Stiewe,
J.\ Phys.\ G {\bf 39}, 083002 (2012).

\bibitem{AnalogueGravity} 
See, e.g., 
C.~Barcel\'o, S.~Liberati, and M.~Visser,
Living Rev.\ Rel.\  {\bf 8}, 12 (2005)
[Living Rev.\ Rel.\  {\bf 14}, 3 (2011)].

\bibitem{PTsymQM} 
See, e.g., 
J.~Schindler {\it et al.},
Phys.\ Rev.\ A {\bf 84}, 040101(R)(2011);
C.M.~Bender, B.K.~Berntson, D.~Parker, and E.~Samuel,
Am.\ J.\ Phys.\  {\bf 81}, 173 (2013).

\bibitem{mh}
M.~Matsumoto,
Tensor, NS {\bf 24}, 29 (1972);
M.~Matsumoto and S.~H\=oj\=o,
Tensor, NS {\bf 32}, 225 (1978).

\bibitem{dk14}
D.~Colladay, 
private communication.

\bibitem{Jackson}
See, e.g., 
J.D.~Jackson, 
{\it Classical Electrodynamics}, 
John Wiley \& Sons, New York, 1998, 
3rd Edition.


\end{thebibliography}
\end{document}